\documentclass[12pt]{article}

\usepackage{amsmath,amsfonts,epsfig,mathrsfs,todonotes,yfonts}
\usepackage[driver=pdftex,top=2.5cm, bottom=2.5cm, left=2.5cm, right=2.5cm]{geometry} 
\numberwithin{equation}{section}

\usepackage{hyperref}
 \hypersetup{
     colorlinks=true,
     linkcolor=black,
     filecolor=blue,
     citecolor=blue,      
     urlcolor=cyan,
     }

\newcommand{\re}[1] {(\ref{#1})}
\newcommand{\pa}{\partial}

\newcommand{\ber}{\begin{eqnarray}}
\newcommand{\eer}[1]{\label{#1}\end{eqnarray}}
\newcommand{\eero}{\end{eqnarray}}

\newcommand{\balg}{\begin{align}}
\newcommand{\ealg}{\end{align}}
\newcommand{\bald}{\begin{aligned}}
\newcommand{\eald}{\end{aligned}}

\newcommand{\beq}{\begin{equation}}
\newcommand{\eeq}{\end{equation}}
\newcommand{\bea}{\begin{eqnarray}}
\newcommand{\eea}{\end{eqnarray}}

\newcommand{\nn}{\nonumber}

\newcommand{\auth}{\large Okan Günel \footnote{email: okan@metu.edu.tr}
and {\large \"O}zg{\"u}r Sar{\i}o\u{g}lu \footnote{email: sarioglu@metu.edu.tr}}

\begin{document}

\begin{center}
{\Large{\bf Conserved charges of the Kerr black hole revisited}}
\vspace{.75cm}

\auth
\end{center}
\vspace{.5cm}
%\vspace{.5cm}
\centerline{{\it \small Department of Physics, Faculty of Arts and Sciences,}}
\centerline{{\it \small Middle East Technical University, 06800, Ankara, Turkey}}

\vspace{1cm}

%\today

\centerline{{\bf Abstract}}
\bigskip

\noindent 
We revisit the Kerr black hole as cast in the Boyer-Lindquist, 
Kerr-Schild and Weyl canonical coordinates, and calculate its
total mass/energy and total angular momentum by using 
linearized gravity along with its background Killing 
isometries. We argue that the integration of the relevant 
gravitational flux does not depend on the geometry of the 
closed and simply connected spatial boundary provided 
it is also piecewise smooth.

\bigskip

\vspace{1cm}

\noindent
Dedicated to the memory of Stanley Deser (March 19, 1931 -- April 21, 2023).

\vspace{10cm}

\small

\renewcommand{\thefootnote}{\arabic{footnote}}
\setcounter{footnote}{0}

\pagebreak
\tableofcontents
\setcounter{page}{2}

\section{Introduction}\label{intro}
We are obviously living in a new era of gravitational physics ever 
since the first successful experimental detection of gravitational 
waves from the collision of two black holes 
\cite{LIGOScientific:2016aoc} and the first image of a black hole 
distinguishing its horizon structure 
\cite{EventHorizonTelescope:2019dse} 
in the way predicted by General Relativity. The gravitational wave 
observations so far are thought to consist of the products of 
collisions of compact astrophysical rotating objects.
Since collisions of stationary black holes seem to form 
a rather large class in this catalogue, it is only natural 
that we heavily rely on the celebrated Kerr solution 
\cite{Kerr:1963ud} in the analysis and understanding of the 
basic physical properties of these collision events. 

The Kerr solution has, of course, been thoroughly studied over 
the years and its basic physical and geometrical properties are 
well understood (See e.g. \cite{Visser:2007fj} and 
\cite{Stephani:2003tm} for a concise introduction and starting 
point.). Specifically the determination of its conserved 
gravitational charges, in particular its total mass/energy and 
total angular momentum, is standard textbook material
\cite{Wald:1984rg}. Here we want to \emph{revisit} 
the particular method based on linearized gravity and the use of 
background Killing isometries \cite{Abbott:1981ff}, that is
in fact closely related to the celebrated Arnowitt-Deser-Misner 
(ADM) formulation \cite{Arnowitt:1962hi} of General Relativity. 
We do this in three separate coordinate systems: the ubiquitous 
Boyer-Lindquist coordinates \cite{Boyer:1966qh}, the Kerr-Schild 
form \cite{KerrSchild} and the lesser-known Weyl canonical 
coordinates \cite{Stephani:2003tm, Jones:2005hj}. Our 
motivation in doing so is foremost to examine how 
the integration of the gravitational flux through ``the 
boundary at infinity" depends on the geometry of the latter. For 
the three poses of the Kerr solution 
studied, ``the cube at infinity" for the Kerr-Schild form and 
``the cylinder at infinity" for the Weyl canonical coordinates 
are certainly not smooth geometries as for ``the sphere at 
infinity" for the Boyer-Lindquist case. As a by-product, we 
also explicitly verify that the whole procedure is indeed background 
gauge invariant. 

The organization of the paper is as follows. In section \!\ref{Concharg},
we start by giving a concise, self-contained description of how
gravitational charges are defined using background Killing isometries.
We explicitly calculate the total mass/energy and the total angular 
momentum of the Kerr black hole cast in the Boyer-Lindquist, 
Kerr-Schild and Weyl canonical coordinates in 
sections \!\ref{KM}, \!\ref{KScoord}, \!\ref{Wcoord}, 
respectively. To our knowledge, the discussion presented
in sections \!\ref{KScoord} and  \!\ref{Wcoord} is novel.
We finally conclude with a brief summary and some remarks in section
\!\ref{conc}. Finally, we explicitly relegate some unwieldy 
formulas needed for the discussion in section \!\ref{Wcoord} to the 
appendix \!\ref{appendweyl}.

\section{Conserved charges in linearized gravity}\label{Concharg}
Let us briefly recapitulate how gravitational charges 
that use background Killing isometries are defined. (Please refer
to \cite{Abbott:1981ff} for details.) 
Let $h_{a b}$ denote the ``deviation" of an asymptotically 
flat metric from its flat ``background" 
$\bar{g}_{a b}$\footnote{Note that this discussion in fact 
holds for asymptotically maximally symmetric spacetimes but 
we simply work with a vanishing cosmological constant in 
what follows.}, i.e. \( h_{a b} \equiv g_{a b} - \bar{g}_{a b} \), 
where $\bar{g}_{a b}$ is the Ricci flat background with curvature
tensors \( \bar{R}_{a b} = 0 = \bar{R} \). It is also assumed that the 
deviation $h_{a b}$ goes to zero ``sufficiently fast" as one 
approaches the background $\bar{g}_{a b}$, which is typically 
located at ``the boundary at infinity". It can be shown 
\cite{Abbott:1981ff} that one can construct a conserved vector 
current \( J^{a} := G_{L}^{a b} \bar{\xi}_{b} \), with
\( \bar{\nabla}_{a} J^{a} = 0 \), out of the linearized 
Einstein tensor $G_{L}^{a b}$ and the background Killing vector 
$\bar{\xi}^{a}$, that are well-defined and smooth in 
the geometry described by $\bar{g}_{a b}$. Then the following 
gives a conserved and background gauge invariant gravitational charge 
\begin{equation}
Q (\bar{\xi}) = \frac{1}{8 \pi} \int_{\Sigma} d^{3}x \, 
\sqrt{|\gamma|} \, n_{a} \, J^{a}  
= \frac{1}{8 \pi} \oint_{\partial \Sigma} d^{2}x \, 
\sqrt{|q|} \, {n}_{[a} \, r_{b]} \, \ell^{a b} \,,
\label{charge}
\end{equation}
where $\ell_{a b}$ is the potential 2-form of the conserved
vector current \( J^{a} := \bar{\nabla}_{b} \ell^{a b} \). 

Here we assume that the background admits a foliation
\beq
\bar{g}_{a b} := - n_a n_b + \gamma_ {a b}
= - n_a n_b + r_a r_b + q_{ab} \,,
\label{fol}
\eeq
where the induced non-degenerate metric $\gamma_{a b}$ 
on the 
$3$-dimensional spacelike hypersurface $\Sigma$ has an
everywhere well-defined timelike normal $n^{a}$ and
\( \gamma := \det{\gamma_{a b}} \). We also 
assume that the hypersurface $\Sigma$ has a boundary 
$\partial \Sigma$ with a metric $q_{ab}$, with
\( q := \det{q_{a b}} \); $n^a$ and the spacelike vector 
$r^a$ are mutually orthogonal unit  vectors to 
$\partial \Sigma$, with \( n^a n_a = -1 \) and \( r^a r_a = 1 \).

For the conventions adopted in this work, one has \cite{Abbott:1981ff}
\beq
\ell^{a b} (\bar{\xi}) := 
\bar{\xi}_{c} \bar{\nabla}^{[a} h^{b]c} 
+ \bar{\xi}^{[b} \bar{\nabla}_{c} h^{a]c} 
+ h^{c [b} \bar{\nabla}_{c} \bar{\xi}^{a]} 
+ \bar{\xi}^{[a} \bar{\nabla}^{b]} h 
+ \frac{1}{2} h \bar{\nabla}^{[a} \bar{\xi}^{b]} \,, 
\label{einchar} 
\eeq
where all raising and lowering of indices are done with respect 
to $\bar{g}_{a b}$, $\bar{\nabla}$ 
indicates the covariant derivative with respect to the 
background metric and the antisymmetrization of indices 
is done with `weight 1' i.e. 
\( A_{[a b]} \equiv \tfrac{1}{2} ( A_{a b} - A_{b a} ) \).

All of these are of course well-known. Our aim here is to 
carefully apply these to the celebrated Kerr metric written in
three separate coordinates, and show how ``delicate" 
the integration of the properly weighted 2-form potential 
$\ell$ on the two-dimensional boundary $\partial \Sigma$ 
\re{charge} can be. For warming up, we start our discussion
by first considering the Kerr black hole in the ubiquitous
Boyer-Lindquist coordinates.

\section{Kerr black hole in Boyer-Lindquist coordinates}\label{KM}
The Kerr black hole \cite{Kerr:1963ud} in Boyer-Lindquist coordinates 
\cite{Boyer:1966qh, Visser:2007fj} reads
\bea
ds^{2} & = & - \left( 1- \frac{2 M r}{\Sigma} \right) dt^{2} 
-\frac{4 a M r \sin^2{\theta}}{\Sigma} \, dt \, d\phi 
+ \frac{\Sigma}{\Delta} \, dr^{2} + \Sigma \, d\theta^2 
\nn \\
& & + \left( r^2 + a^2 + \frac{2 a^2 M r \sin^2{\theta}}{\Sigma} \right) \sin^2{\theta} \, d\phi^{2} \,,
\label{KerrBL} 
\eea
where the metric functions are given by
\beq
\Sigma(r, \theta) := r^{2} + a^{2} \cos^{2}{\theta} 
\qquad \mbox{and} \qquad
\Delta(r) := r^{2} + a^{2} - 2 M r \,. \label{TheSigDel}
\eeq
The relevant flat background is found by setting 
\( M=0, a=0 \) in \re{KerrBL}:
\beq
\label{KNback}
\bar{g}_{ab} = {\rm diag} (-1, 1, r^2, r^2 \sin^{2}\theta) \,,
\eeq
which admits the foliation \re{fol} with
\beq
\bald
& \gamma_{a b} = {\rm diag} (0, 1, r^2, r^2 \sin^{2}\theta) \quad \mbox{and} \quad
q_{ab}={\rm diag} (0, 0, r^2, r^2 \sin^{2}\theta) \,, \\
& n^a = (-\pa_{t})^a \quad \mbox{and} \quad 
r^a = (\pa_{r})^a \,.
%\;\; \mbox{with} \;\; 
%n^a n_a = -1 
%\;\; \mbox{and} \;\; 
%r^a r_a = 1 \,,
\eald
\eeq
Clearly $\gamma_{ab}$ is the metric on the 
3-dimensional Euclidean space $\Sigma$ in spherical 
polar coordinates and $q_{ab}$ is the metric on 
a 2-sphere $S_2$ of radius $r$. The 2-dimensional boundary 
$\pa \Sigma$ is found when one takes
$r \to \infty$. The background clearly has 
two globally defined, timelike  
$\bar{\xi}^{a} = \left( - \pa_t \right)^{a}$
and spacelike 
$\bar{\zeta}^{a} = \left( \pa_{\phi} \right)^{a}$,
Killing vectors.

The relevant components of the 2-form 
potential \re{einchar} for each Killing vector are
\begin{align}
& \ell^{t r} (\bar{\xi}) = \frac{1}{2 r^3} \left(
\frac{2 a^2 M r^2 \sin^2{\theta} \Sigma'}{\Sigma^2} 
+ \left( \frac{2 r^2}{\Delta} + 1 \right) \Sigma 
- r \Sigma' + a^2 - r^2 \right)
\,, \\
& \ell^{t r} (\bar{\zeta}) =
\frac{a M \sin^2{\theta} \left( r \Sigma' + \Sigma \right)}{\Sigma^2} \,,
\end{align}
in terms of the metric functions \re{TheSigDel} where
a prime indicates derivative with 
respect to the $r$ coordinate. A careful integration
gives
\begin{equation}
\oint_{S_2} r^2 d\Omega \,
{n}_{[a} \, r_{b]} \, \ell^{a b} (\bar{\xi}) =
\frac{4 \pi}{3} \left(\frac{r \left(a^2+3 r^2\right)}{\Delta}+\frac{2 a^3+3 M \left(a^2-r^2\right) 
\arctan{(a/r)} +3 a r (M-r)}{a r} \right),
\end{equation}
where we have used 
\( d\Omega := \sin{\theta} \, d\theta \, d\phi \),
that leads to
\ber
Q(\bar{\xi}) = \frac{1}{8 \pi} \lim_{r \to \infty} 
\oint_{S_2} r^2 d\Omega \, {n}_{[a} \, r_{b]} \, \ell^{a b} (\bar{\xi}) = M \,,
\eer{MassBL}
the total mass of the Kerr black hole. Similarly, one finds
\begin{equation}
\oint_{S_2} r^2 d\Omega \,
{n}_{[a} \, r_{b]} \, \ell^{a b} (\bar{\zeta}) =
8 \pi M r \arctan{(a/r)},
\end{equation}
which yields
\ber
Q(\bar{\zeta})= \frac{1}{8 \pi} \lim_{r \to \infty} 
\oint_{S_2} r^2 d\Omega \, {n}_{[a} \, r_{b]} 
\, \ell^{a b} (\bar{\zeta}) = a M \,,
\eer{AngMomBL}
the total angular momentum of the Kerr black hole.

At hindsight, what we did was to first integrate the properly furbished
2-form potential $\ell$ on a constant radius 2-sphere $S_2$ and 
later take the radius $r \to \infty$ limit. It is perhaps worthy of
considering the ``short-cut" of taking the $r \to \infty$ limit
of the properly furbished 2-form potential $\ell$ first and then
integrate the result on ``the sphere at infinity" later. This 
interchange is in general clearly not up to par with mathematical 
rigor, but it should be possible for a ``smooth" boundary 
$\pa \Sigma$ and a ``smooth" potential $\ell$. Thus, e.g.
\begin{align}
& \frac{1}{8 \pi} \oint d\Omega \left( \lim_{r \to \infty} 
r^2 \, {n}_{[a} \, r_{b]} \, \ell^{a b} (\bar{\xi}) \right)
= \frac{1}{8 \pi} \oint d\Omega \, (2 M) = M \,, \nn \\
& \frac{1}{8 \pi} \oint d\Omega \left( \lim_{r \to \infty} 
r^2 \, {n}_{[a} \, r_{b]} \, \ell^{a b} (\bar{\zeta}) \right)
= \frac{1}{8 \pi} \oint d\Omega \, (3 a M \sin^{2}{\theta}) = a M \,,
\end{align}
which ``validates" our hunch for the shortcut. 

\section{Kerr black hole in Kerr-Schild coordinates}\label{KScoord}
Let us consider the Kerr black hole in its Kerr-Schild form \cite{KerrSchild,Myers:1986un,Gibbons:2004uw}:
\ber
d s^{2} = g_{ab} d x^{a} d x^{b}
= \left( \eta_{ab}  + \frac{2 M}{U} \lambda_{a} \lambda_{b} \right) d x^{a} d x^{b} \,,
\eer{KS}
where $\lambda^{a}$ is null and geodesic with respect to both 
the full metric $g_{ab}$ and the flat Minkowski metric $\eta_{ab}$. 
Explicitly, in flat coordinates $(t, x, y, z)$, one has
\ber
\lambda = \lambda_{a} d x^{a} = d t + \frac{ r (x d x + y d y) + a (y d x - x d y)}{r^{2} + a^{2}} 
+ \frac{z d z}{r}
\eer{lam}
and
\ber
U = r + \frac{a^{2} z^{2}}{r^{3}} \,,
\eer{UU}
where $r$ is not a coordinate but a function of $(x,y,z)$ and defined by
\ber
\frac{x^{2} + y^{2}}{r^{2} + a^{2}} + \frac{z^{2}}{r^{2}} = 1 \,.
\eer{rdef}
Note that the Eddington-like coordinates \cite{Boyer:1966qh}
$(\bar t,r,\theta, \bar \phi)$  can be obtained from \re{KS} 
by the coordinate transformation
\bea
t &=& \bar{t}  \,. \nn \\
x &=& \sqrt{r^2 +a^2} \, \sin{\theta} \, 
\cos{ \left( \bar{\phi} + \arctan{(a/r)} \right)} \,, \nn \\
y &=& \sqrt{r^2 +a^2} \, \sin{\theta} \, 
\sin{\left( \bar{\phi} + \arctan{(a/r)} \right)} \,, \label{CoTr} \\
z &=& r \cos{\theta} \,, \nn 
\eea
A final time-dependent coordinate transformation is required to arrive at the 
Kerr metric in Boyer-Lindquist coordinates $(t, r, \theta, \phi)$ as in 
\re{KerrBL}
\begin{eqnarray*}
d\bar{t} &=& dt + \frac{2 M r}{\Delta} \, dr \,,  \\
d\bar{\phi} &=& d\phi + \frac{a}{\Delta} \, dr \,. 
\end{eqnarray*}

One advantage of the Kerr-Schild form \re{KS} is that 
the background metric is clearly the flat Minkowski metric 
$\eta_{ab}$ in the usual Cartesian coordinates. This background 
admits the foliation \re{fol} with
\ber
\gamma_{a b} = {\rm diag} (0, 1, 1, 1) \,,
\quad n^a = (-\pa_{t})^a \,, 
\eer{gammnup}
where $\gamma_{a b}$ is the metric on the 3-dimensional 
flat Euclidean space in the Cartesian coordinates.
Now arises the question about the integration on the boundary 
of the space. The boundary relevant for the charge
definition \re{charge} can be thought of as ``the cube at
infinity" in Cartesian coordinates. This can be described
best by first considering a cube centered at the origin and
of side length $2L$, and then taking the $L \to \infty$ limit.

Any such cube of finite side length has six faces and the 
integration on each face must be carried out with the 
following normal vectors and the corresponding 
two-dimensional metrics in mind:
\begin{align}
& x_{\pm}^{a} = \big( \pm \pa_x \big)^{a} 
\quad \mbox{and} \quad
\big( q_x \big)_{ab} = {\rm diag} (0, 0, 1, 1) \,, \nn \\ 
& y_{\pm}^{a} = \big( \pm \pa_y \big)^{a} 
\quad \mbox{and} \quad
\big( q_y \big)_{ab} = {\rm diag} (0, 1, 0, 1) \,, \label{xuyuzu} \\
& z_{\pm}^{a} = \big( \pm \pa_z \big)^{a}  
\quad \mbox{and} \quad
\big( q_z \big)_{ab} = {\rm diag} (0, 1, 1, 0) \,. \nn
\end{align}
``The cube at infinity" presumably has its faces placed 
at the $L \to \infty$ limits of the coordinates along the 
direction of the normal vectors. Thus the integration on the 
boundary can be thought of as the sum of six separate 
integrals on each face of ``the cube at infinity"
\begin{align}
Q (\bar{\xi}) = & \frac{1}{8 \pi} \lim_{L \to \infty}
\left( 
\int_{-L}^{L} \int_{-L}^{L} dy \, dz \, 
{n}_{[a} \, x_{b]} \, \ell^{a b}  (\bar{\xi}) \rvert_{x=L}
-\int_{-L}^{L} \int_{-L}^{L} dy \, dz \, 
{n}_{[a} \, x_{b]} \, \ell^{a b}  (\bar{\xi}) \rvert_{x=-L} \right. \nn \\ & \qquad \qquad
+ \int_{-L}^{L} \int_{-L}^{L} dx \, dz \, 
{n}_{[a} \, y_{b]} \, \ell^{a b} (\bar{\xi}) \rvert_{y=L}
- \int_{-L}^{L} \int_{-L}^{L} dx \, dz \, 
{n}_{[a} \, y_{b]} \, \ell^{a b} (\bar{\xi}) \rvert_{y=-L}
 \nn \\ 
& \qquad \qquad \left.
+ \int_{-L}^{L} \int_{-L}^{L} dx \, dy \, 
{n}_{[a} \, z_{b]} \, \ell^{a b} (\bar{\xi}) \rvert_{z=L} 
- \int_{-L}^{L} \int_{-L}^{L} dx \, dy \, 
{n}_{[a} \, z_{b]} \, \ell^{a b} (\bar{\xi}) \rvert_{z=-L} 
\right) \,. 
\end{align} 
Using \re{gammnup} and \re{xuyuzu}, this explicitly becomes
\begin{align}
\label{chargeatcube}
Q (\bar{\xi}) = & 
\frac{1}{8 \pi} \lim_{L \to \infty} \left(
\int_{-L}^{L} \int_{-L}^{L} dy \, dz \, \ell^{t x}(\bar{\xi}) \rvert_{x=L} - 
\int_{-L}^{L} \int_{-L}^{L} dy \, dz \, \ell^{t x}(\bar{\xi}) \rvert_{x=-L}
\right. \nn \\
& \qquad \qquad 
+ \int_{-L}^{L} \int_{-L}^{L} dx \, dz \, \ell^{t y}(\bar{\xi}) \rvert_{y=L}
- \int_{-L}^{L} \int_{-L}^{L} dx \, dz \, \ell^{t y}(\bar{\xi}) \rvert_{y=-L} \nn \\ 
& \qquad \qquad \left.
+ \int_{-L}^{L} \int_{-L}^{L} dx \, dy  \, \ell^{t z}(\bar{\xi}) \rvert_{z=L}
- \int_{-L}^{L} \int_{-L}^{L} dx \, dy  \, \ell^{t z}(\bar{\xi}) \rvert_{z=-L}
\right) \,,
\end{align} 
keeping $a$ and $M$ fixed, and the definition of $r$ \re{rdef} in mind.
The relevant timelike Killing vector of the background is 
still $\bar{\xi}^{a} = \big( -\pa_t \big)^{a}$,
which generates three nontrivial components for 
the 2-form potential $\ell$ \re{einchar}
\begin{subequations}\label{potKS}
\begin{align}
\ell^{tx} (\bar{\xi}) = & 
\frac{M r^2}{(a^2+r^2)^3 (a^2 z^2+r^4)^3}
\Big( 2 r^{13} x + a^9 y z^4-2 a^8 r x z^4-2 a^7 r^2 y z^2 (r^2-3 z^2) \Big. \nn \\
& \quad -a^6 r^3 x z^2 \big( 4 r^2+3 (x^2+y^2) \big)
+a^5 r^4 y \big( r^4+z^2 (4 x^2+4 y^2+5 z^2) \big)
\nn \\
& \quad
-a^4 r^5 x (r^2+z^2) (2 r^2-x^2-y^2-2 z^2)
+2 a^3 r^8 y (3 r^2-z^2) \\
& \quad \Big. +a^2 r^9 x (5 x^2+5 y^2+4 z^2)
+a r^{10} y \big( 5 r^2-4 (x^2+y^2+z^2) \big)
\Big) \,, \nn \\
\ell^{ty}(\bar{\xi}) = & 
\frac{M r^2}{(a^2+r^2)^3 (a^2 z^2+r^4)^3}
\Big( 2 r^{13} y - a^9 x z^4-2 a^8 r y z^4+2 a^7 r^2 x z^2 (r^2-3 z^2) \Big. \nn \\
& \quad -a^6 r^3 y z^2 \big(4 r^2+3 (x^2+y^2) \big)
-a^5 r^4 x \big( r^4+z^2 (4 x^2+4 y^2+5 z^2) \big)
\nn \\
& \quad
-a^4 r^5 y (r^2+z^2) (2 r^2-x^2-y^2-2 z^2)
+2 a^3 r^8 x (z^2-3 r^2) \\
& \quad \Big. +a^2 r^9 y (5 x^2+5 y^2+4 z^2)
+a r^{10} x \big(4 (x^2+y^2+z^2)-5 r^2 \big)
\Big) \,, \nn \\
\ell^{tz}(\bar{\xi}) = & 
\frac{M r^3 z}{(a^2+r^2)^2 (a^2 z^2+r^4)^3} 
\Big( 2 r^{10} -a^6 z^2 (x^2+y^2-2 z^2)
\Big. \\
& \quad \Big. +a^4 \big(r^4 (3 x^2+3 y^2+4 z^2)
+r^2 z^2 (x^2+y^2+2 z^2) \big)
+a^2 r^6 (2 r^2+5 x^2+5 y^2+4 z^2) \Big) \,. \nn
\end{align}
\end{subequations}
To arrive at the expressions \re{potKS}, we have utilized
\begin{align}
& \frac{\pa r}{\pa x} = \frac{x}{U} \,, \qquad
\frac{\pa r}{\pa y} = \frac{y}{U} \,, \qquad
\frac{\pa r}{\pa z} = \frac{z}{U} \Big( 1+ \frac{a^2}{r^2} \Big) \,, \label{parr} \\
& \frac{\pa U}{\pa x} = \frac{x}{U} \Big( 1- \frac{3 a^2 z^2}{r^4} \Big) \,, \quad
\frac{\pa U}{\pa y} = \frac{y}{U} \Big( 1- \frac{3 a^2 z^2}{r^4} \Big) \,, \quad
\frac{\pa U}{\pa z} = \frac{z}{U} \Big( 1- \frac{3 a^2 z^2}{r^4} \Big) \Big( 1+ \frac{a^2}{r^2} \Big) + \frac{2 a^2 z}{r^3} \,,
\label{parU}
\end{align}
that follow from the definitions of $r$ \re{rdef} and 
$U$ \re{UU}, respectively.

Next we substitute \re{potKS} into \re{chargeatcube}. Notice that 
the variable $r$ \re{rdef} is symmetric in $x$ and $y$ by 
definition. Thus,  we can interchange the latter in the 
integration for the terms that include, e.g., $\ell^{t y}$. A 
careful consideration along the $yz$- and $xz$-surfaces leads to 
\begin{align}
Q(\bar{\xi}) =  & \frac{M}{4 \pi}  \lim_{L \to \infty} \bigg(
\int_{-L}^{L} \int_{-L}^{L} dx \, dz \, 
\frac{2 \, r(x,L,z)^{3} L}{(r^2 + a^2)^3 (r^4 + a^2 z^2)^3} \left(
 2 r^{12} -2 a^8 z^4 - a^6 r^2 z^2 \big( 4 r^2+3 (x^2+L^2) \big) \right. \nn \\
 & \qquad \qquad \qquad \left.
-a^4 r^4 (r^2+z^2) (2 r^2-x^2-L^2-2 z^2)
 +a^2 r^8  (5 x^2+5 L^2+4 z^2)
\right) \nn \\ 
& + \int_{-L}^{L} \int_{-L}^{L} dx \, dy \, 
\frac{ r(x,y,L)^{3} L}{(r^2 + a^2)^2 (r^4 + a^2 L^2)^3} \left(
 2 r^{10} - a^6 L^2  (x^2+y^2 - 2 L^2) \right. \label{ara1} \\
 & \qquad \left.
+ a^4 r^2 \big( r^2 (3 x^2 + 3 y^2 + 4 L^2) + L^2 (x^2+y^2+2L^2) \big) 
 +a^2 r^6  (2 r^2 + 5 x^2+5 y^2+4 L^2)
\right)
\bigg) \,, \nn
\end{align}
where we have implicitly indicated the relevant $y=\pm L$ and 
$z=\pm L$ substitutions in $r$ \re{rdef} by writing it with
its pertinent arguments. Using \re{rdef} to eliminate $x^2$ and
$x^2+y^2$ combinations in the two integrands of \re{ara1}, 
respectively, we arrive at
\begin{align}
Q (\bar{\xi}) =  & \frac{M}{4 \pi}  \lim_{L \to \infty} \bigg(
\int_{-L}^{L} \int_{-L}^{L} dx \, dz \, 
\frac{2 \, r(x,L,z)^{3} L}{(r^2 + a^2) (r^4 + a^2 z^2)^3} \left(
 2 r^{8} +a^{4} z^{2} (z^{2}-3r^{2}) +a^{2} r^{4} (r^{2}-z^{2})
\right) \nn \\ 
& \qquad \qquad   + \int_{-L}^{L} \int_{-L}^{L} dx \, dy \, 
\frac{ r(x,y,L) L }{\left(r^4+a^2 L^2\right)^3}
\left(2 r^8 - a^4 L^2 ( r^2-L^2) - a^2 r^4 ( L^2-3 r^2)\right)
\bigg) \, \label{ara2}
\end{align}
where the first integrand can be factorized to write
\begin{align}
Q (\bar{\xi}) & = I_{xy} + I_{xz} \,, \label{IxyIxz} \\
I_{xy} & := \frac{M}{4 \pi}  \lim_{L \to \infty} \bigg(
\int_{-L}^{L} \int_{-L}^{L} dx \, dy \, 
\frac{ r(x,y,L) L }{\left(r^4+a^2 L^2\right)^3}
\left(2 r^8 - a^4 L^2 ( r^2-L^2) - a^2 r^4 ( L^2-3 r^2)\right)
\bigg) \,, \nn \\
I_{xz} & := \frac{M}{4 \pi}  \lim_{L \to \infty} \bigg(
\int_{-L}^{L} \int_{-L}^{L} dx \, dz \, 
\frac{2 \, r(x,L,z)^{3} L}{(r^4 + a^2 z^2)}
\left( \frac{r^{2} (r^4-3a^2 z^2)}{(r^4 + a^2 z^2)^2}+
\frac{1}{r^2+a^2} \right) \nn \,.
\end{align}
Since the solution of $r$ \re{rdef} is not as simple as in the usual 
spherical coordinates, it is painful to take these integrals in the 
Euclidean coordinates. 

For dealing with $I_{xy}$, we first note that the relevant
root of $r=r(x,y,L)$ \re{rdef} is
\ber
r = \frac{\sqrt{\sqrt{\left(-a^2+L^2+x^2+y^2\right)^2
+4 a^2 L^2}-a^2+L^2+x^2+y^2}}{\sqrt{2}} \,.
\eer{rxy}
This is to be substituted into $I_{xy}$ in \re{IxyIxz}.
Note, however, that the resultant expression is an analytic function of 
the rotation parameter $a$ and that we are to first do the 
integration on the $x$ and $z$ variables, then set 
$x = \pm L$ and $z = \pm L$ and next take the $L \to \infty$
limit, keeping the parameter $a$ finite and fixed the whole time. 
Thus the whole integral $I_{xy}$ must be an analytic function of the 
parameter $a$, or better yet, $u := a/L$ about $u=0$ for fixed $a$. 
Specifically, we find 
\[ \frac{2 L}{\left(L^2+x^2+y^2\right)^{3/2}}
+\frac{2 u^2 L^3 \left(3 \left(x^2+y^2\right)-2 L^2\right)}
{\left(L^2+x^2+y^2\right)^{7/2}} 
+\frac{3 u^4 L^5 \left(8 L^4-40 L^2 
\left(x^2+y^2\right)+15 
\left(x^2+y^2\right)^2\right)}{4 
\left(L^2+x^2+y^2\right)^{11/2}} \]
for the integrand of $I_{xy}$ to quintic order 
about $u=0$. We next integrate this $u$-series
term by term in both $x$ and $y$ over the square
$x \in (-L,L)$ and $y \in (-L,L)$ to arrive at
\ber
I_{xy} = \frac{M}{108 \pi} \, \lim_{u \to 0} \,
\left( 36 \pi + \sqrt{3} 
\left(-32+ u^2\right) u^2 \right) 
= \frac{M}{3} \,.
\eer{Ixyson} 

As for \( I_{xz} \), we similarly have
\ber
r= \frac{\sqrt{\sqrt{\left(-a^2+L^2+x^2+z^2\right)^2+4 a^2 z^2}-a^2+L^2+x^2+z^2}}{\sqrt{2}}
\eer{rxz}
for the root of $r=r(x,L,z)$ \re{rdef}, and the corresponding
series expansion in $u$ is
\[ \frac{4 L}{\left(L^2+x^2+z^2\right)^{3/2}}
+\frac{4 u^2 L^3 \left(x^2+L^2-4 z^2\right)}
{\left(L^2+x^2+z^2\right)^{7/2}} 
+\frac{9 u^4 L^5 \left(8 z^4-12 z^2 
\left(x^2+L^2\right)+ 
\left(x^2+L^2\right)^2\right)}{2 
\left(L^2+x^2+z^2\right)^{11/2}} \]
for the integrand of $I_{xz}$ to quintic order
about $u=0$. We integrate term by term in both $x$ and $z$ over the square 
$x \in (-L,L)$ and $z \in (-L,L)$, and get
\ber
I_{xz} = \frac{M}{108 \pi} \, \lim_{u \to 0} \,
\left( 72 \pi + \sqrt{3} 
\left(32- u^2\right) u^2 \right) 
= \frac{2 M}{3} \,.
\eer{Ixzson}
Hence the total mass \( Q(\bar{\xi}) \) is
\ber
Q(\bar{\xi}) = M \,,
\eer{MassKS}
in accordance with \re{MassBL}.

The azimuthal (rotational) Killing vector reads
\ber
\bar{\zeta}^a = \left( x \, \pa{_y} - y \, \pa_{x} \right)^{a} \,.
\eer{rotKV}
in the Kerr-Schild coordinates. Using the same foliation \re{gammnup}
and the six boundary faces \re{xuyuzu}, we end up with
\( Q(\bar{\zeta}) \) similar to \re{chargeatcube}. This time the
three nontrivial components of the 2-form potential $\ell$ \re{einchar}
are
\begin{subequations}\label{potKSrot}
\begin{align}
\ell^{tx} (\bar{\zeta}) = &
\frac{M r^2}{(a^2+r^2)^3 (a^2 z^2+r^4)^3}
\Big( a^8 y z^4-2 a^6 r^2 y z^2 \left(r^2-2 z^2 \right)
- 3 a^5 r^3 x z^2 \left(x^2+y^2\right) 
\Big. \nn \\
& \quad 
+a^4 r^4 y \left(r^4-6 r^2 z^2+z^2 \left(4 x^2+4 y^2+5 z^2\right)\right)
+a^3 r^5 x \left(r^2-z^2\right) \left(x^2+y^2\right)
\Big. \nn \\
& \quad \Big.
+2 a^2 r^6 y \left(r^4-3 r^2 z^2+z^2 \left(x^2+y^2+z^2\right)\right)
+3 a r^9 x \left(x^2+y^2\right)
\Big. \\
& \quad +r^{10} y \left(r^2-2 \left(x^2+y^2+z^2\right)\right)
\Big) \,, \nn \\
\ell^{ty}(\bar{\zeta}) = & 
- \frac{M r^2}{(a^2+r^2)^3 (a^2 z^2+r^4)^3}
\Big( a^8 x z^4-2 a^6 r^2 x z^2 \left(r^2-2 z^2\right)
+ 3 a^5 r^3 y z^2 \left(x^2+y^2\right)
\Big. \nn \\
& \quad 
+a^4 r^4 x \left(r^4-6 r^2 z^2+z^2 \left(4 x^2+4 y^2+5z^2\right)\right)
-a^3 r^5 y \left(r^2-z^2\right) \left(x^2+y^2\right)
\Big. \nn \\
& \quad \Big.
+2 a^2 r^6 x \left(r^4-3 r^2 z^2+z^2 \left(x^2+y^2+z^2\right)\right)
-3 a r^9 y \left(x^2+y^2\right)
\Big. \\
& \quad +r^{10} x \left(r^2-2 \left(x^2+y^2+z^2\right)\right)
\Big) \,, \nn \\
\ell^{tz}(\bar{\zeta}) = & 
 -\frac{a M r^3 (x^2 + y^2) z (-3 r^4 + a^2 z^2)}{(r^4 + a^2 z^2)^3}
 \,, 
\end{align}
\end{subequations}
where we have again used \re{parr} and \re{parU}. We next
substitute \re{potKSrot} into \( Q(\bar{\zeta}) \). Once again 
using the symmetry \( r(x,y) = r(y,x)\) \re{rdef}, 
we can add the integrals on the $xz$- and $yz$-surfaces (in
complete analogy to how \re{IxyIxz} was found) to finally arrive at
\begin{align}
Q(\bar{\zeta}) & = \frac{a M}{4 \pi} \left( J_{xy} + J_{xz} \right) \,, \label{JxyJxz} \\
J_{xy} & :=   \lim_{L \to \infty} \bigg(
\int_{-L}^{L} \int_{-L}^{L} dx \, dy \, 
\frac{ r(x,y,L) L \left(r^2+ a^2\right) \left(r^2-L^2\right) \left(3 r^4-a^2 L^2\right)}{\left(r^4+a^2 L^2\right)^3}
\bigg) \,, \nn \\
J_{xz} & :=   \lim_{L \to \infty} \bigg(
\int_{-L}^{L} \int_{-L}^{L} dx \, dz \, 
\frac{2 \, r(x,L,z)^3 L \left(r^2-z^2\right) \left(3 r^6 + a^2 r^4 - 3 a^4 z^2-a^2 r^2 z^2\right)}{\left(r^2+ a^2\right) \left(r^4+ a^2 z^2\right)^3} \bigg) \nn \,. 
\end{align}

We calculate $J_{xy}$ similar to how we have calculated $I_{xy}$.
After the substitution of \re{rxy}, the series expansion of the 
integrand of $J_{xy}$ about $u=0$ is
\begin{align*}
& \frac{3 L \left(x^2+y^2\right)}{\left(L^2+x^2+y^2\right)^{5/2}}-\frac{5 \left(L^3 \left(x^2+y^2\right) \left(4 L^2-3 \left(x^2+y^2\right)\right)\right)}{2 \left(L^2+x^2+y^2\right)^{9/2}} u^2
\\
& +\frac{21 L^5 \left(x^2+y^2\right) \left(8 L^4-20 L^2 \left(x^2+y^2\right)+5 \left(x^2+y^2\right)^2\right)}{8 \left(L^2+x^2+y^2\right)^{13/2}} u^4 + O(u^5) \,.
\end{align*}
Similarly, the substitution of \re{rxz}
gives the series expansion of the integrand of $J_{xz}$ as
\begin{align*}
& \frac{6 L \left(L^2+x^2\right)}{\left(L^2+x^2+z^2\right)^{5/2}}
+\frac{5 L^3 \left(L^2+x^2\right) \left(L^2+x^2-6 z^2\right)}{\left(L^2+x^2+z^2\right)^{9/2}} u^2 
\\
& +\frac{21 L^5 \left(L^2+x^2\right) \left(-16 z^2 \left(L^2+x^2\right)+\left(L^2+x^2\right)^2+16 z^4\right)}{4 \left(L^2+x^2+z^2\right)^{13/2}} u^4 + O(u^5) \,.
\end{align*}
The integration of these on the relevant squares then leads to
\begin{align}
& J_{xy} = \frac{a M}{2430 \pi} \, \lim_{u \to 0} \,
\left(810 (\pi - \sqrt{3}) -495 \sqrt{3} u^2 + 82 \sqrt{3} u^4 \right) 
= \frac{a M}{3}-\frac{a M \sqrt{3}}{3 \pi} \,, \label{jxy} \\
& J_{xz} = \frac{a M}{2430 \pi} \, \lim_{u \to 0} \,
\left( 810 (2 \pi + \sqrt{3}) +495 \sqrt{3} u^2 - 82 \sqrt{3} u^4 \right)
= \frac{2 a M}{3} + \frac{a M \sqrt{3}}{3 \pi}  \,. \label{jxz}
\end{align}
giving the total angular momentum
\ber
Q(\bar{\zeta}) = a M \,,
\eer{AngMomKS}
in accordance with \re{AngMomBL}.

\section{Kerr black hole in Weyl canonical coordinates}\label{Wcoord}
The Kerr black hole can also be written in the Weyl canonical
coordinates \cite{Stephani:2003tm,Jones:2005hj}
\ber
ds^2 = - e^{2U} ( dt + A \, d\phi )^{2} + e^{-2U} \left( e^{2 k} \left( d\rho^{2} + dz^2 \right) 
+ \rho^{2} d\phi^{2} \right) \,,
\eer{Weyl}
where
\begin{align}
& e^{2U} = \frac{\left(R_{+}+R_{-}\right)^{2}-4 M^{2}+\frac{a^{2}}{M^{2}-a^{2}}\left(R_{+}-R_{-}\right)^{2}}{\left(R_{+}+R_{-}+2 M\right)^{2}+\frac{a^{2}}{M^{2}-a^{2}}\left(R_{+}-R_{-}\right)^{2}} \,,  \\
& e^{2k} = \frac{\left(R_{+}+R_{-}\right)^{2}-4 M^{2}+\frac{a^{2}}{M^{2}-a^{2}}\left(R_{+}-R_{-}\right)^{2}}{4 R_{+} R_{-}} \,,  \\
& A = \frac{2 a M\left(M+\frac{R_{+}+R_{-}}{2}\right)\left(1-\frac{\left(R_{+}-R_{-}\right)^{2}}{4\left(M^{2}-a^{2}\right)}\right)}{\frac{1}{4}\left(R_{+}+R_{-}\right)^{2}-M^{2}
+ a^{2} \frac{\left(R_{+}-R_{-}\right)^{2}}{4\left(M^{2}-a^{2}\right)}} \,,  \\
& R_{\pm} = \sqrt{\rho^{2}+\left(z \pm \sqrt{M^{2}-a^{2}}\right)^{2}}
= r-M \pm \sqrt{M^{2}-a^{2}} \, \cos{\theta} \,. 
\end{align}
The transformation back to Boyer-Lindquist coordinates 
\( (t, r, \theta, \phi )\) \re{KerrBL} is given by
\ber
\rho = \sqrt{r^{2}-2 M r+a^{2}} \, \sin{\theta} \,, \qquad 
z = (r-M) \cos{\theta} \,.
\eer{BLtoW}

The flat background can be obtained at the limit of $M=0,\, a=0$ in \re{Weyl}: 
\ber
\bar{g}_{ab} = {\rm diag} (-1, 1, \rho^{2}, 1) \,,
\eer{Weylback}
and it admits the foliation \re{fol} with
\beq
\gamma_{ab}= {\rm diag} (0, 1, \rho^{2}, 1) \,, \quad 
n^{a}= \left( - \pa_{t} \right)^{a} \,.
\eeq
$\gamma_{ab}$ is the metric on the 3-dimensional Euclidean space 
in cylindrical polar coordinates. For this particular case, let us first 
consider a right-angled circular cylinder which has its axis 
of symmetry on the $z$-axis, of radius $R$ and of height $2L$. 
There are three faces in total, the two top and bottom caps/disks 
and the side surface, with the following normal vectors and the 
corresponding two-dimensional metrics, respectively:
\begin{align}
& z_{\pm}^{a} = \left( \pm \pa_{z} \right)^{a} \quad \mbox{and} \quad
\big( q_{z} \big)_{ab} = {\rm diag} \big(0, 1 , \rho^{2}, 0 \big) \,, \\
& \rho^{a} = \left( \pa_{\rho} \right)^{a} \quad \mbox{and} \quad
\big( q_{\rho} \big)_{ab} = {\rm diag} \big( 0, 0, \rho^{2}, 1 \big)\,.
\end{align}
The boundary is then ``the cylinder at infinity" found by taking the
$R \to \infty$ and $L \to \infty$ limits simultaneously along the
directions of the normal vectors. Finally, the background again has 
two globally defined, timelike  
\( \bar{\xi}^{a} = \left( - \pa_{t} \right)^{a} \)
and spacelike 
\( \bar{\zeta}^{a} = \left( \pa_{\phi} \right)^{a} \), Killing vectors, and 
the integration on the boundary amounts to three separate
integrals on the three faces of ``the cylinder at infinity"
\begin{align}
Q(\bar{\xi}) & = \frac{1}{8 \pi} 
\lim_{L \to \infty} \bigg(
\int_{0}^{2 \pi} \int_{0}^{R} d \rho \, d \phi \, \rho \, 
{n}_{[a} \, z_{+ \, b]} \ell^{ab} (\bar{\xi}) \rvert_{z=L} 
+ \int_{0}^{2 \pi} \int_{0}^{R} d \rho \, d \phi \, \rho \, 
{n}_{[a} \, z_{- \, b]} \ell^{ab} (\bar{\xi}) \rvert_{z=-L} 
\bigg) \nn \\ & \qquad \quad 
+ \frac{1}{8 \pi}  
\lim_{R \to \infty} 
\int_{0}^{2 \pi} \int_{-L}^{L} dz \, d \phi \, \rho \, 
{n}_{[a} \, \rho_{b]} \ell^{a b} (\bar{\xi}) \rvert_{\rho=R} \,, 
\nn \\
& = \frac{1}{4} 
\lim_{\substack{L \to \infty \\ R \to \infty}}
\int_{0}^{R} d \rho \, \rho \Big(
 \ell^{tz} (\bar{\xi}) \rvert_{z=L}
- \ell^{tz} (\bar{\xi}) \rvert_{z=-L}
\Big) + \frac{1}{4} 
\lim_{\substack{R \to \infty \\ L \to \infty}}
\int_{-L}^{L} dz \, R \, \ell^{t\rho} (\bar{\xi}) \rvert_{\rho=R}
\,,
\end{align}
where $a$ and $M$ is to be kept fixed, and we used the fact that 
all relevant parts are independent of $\phi$. We relegate the 
explicit 
and unsavory forms of the two nontrivial components of the 
2-form potential $\ell$ \re{einchar} for the two Killing vectors 
$\bar{\xi}$ and $\bar{\zeta}$ to appendix \ref{appendweyl}. We 
calculate the integrals separately as we did in section \!\ref{KScoord} 
by defining
\begin{align}
Q(\bar{\xi}) & = I_{\rho} + I_{z} \,, \label{IrhoIz} \\
I_{\rho} & :=\frac{1}{4} 
\lim_{\substack{L \to \infty \\ R \to \infty}}
\int_{0}^{R} d \rho \, \rho \Big(
 \ell^{tz} (\bar{\xi}) \rvert_{z=L}
- \ell^{tz} (\bar{\xi}) \rvert_{z=-L}
\Big) \,, \nn \\
I_{z} & := \frac{1}{4} \lim _{\substack{R \to \infty \\ L \to \infty}}
\int_{-L}^{L} dz \, R \, \ell^{t\rho} (\bar{\xi}) \rvert_{\rho=R} \nn \,.
\end{align}

In Kerr-Schild coordinates, the integrands of $I_{xy}$, $I_{xz}$ \re{IxyIxz}
and $J_{xy}$, $J_{xz}$ \re{JxyJxz} depended only on the rotation parameter $a$,
and the mass parameter $M$ came out as an overall factor.  
However, as can be seen from the relevant components of the 
2-form potential $\ell$ in appendix \ref{appendweyl}, the integrands 
of $I_{\rho}$ and $I_{z}$ \re{IrhoIz} explicitly depend on 
both $a$ and $M$. Moreover, there is also the complication that arises from 
taking the two separate limits \( R \to \infty \) and \( L \to \infty\). 
In order to have better control on the geometry and these limits, we set 
\( L = k R \), where $k>0$ is a \emph{finite chubbiness parameter} between 
the length $L$ and the radius $R$ of ``the cylinder at infinity", just before the limiting step and take \( R \to \infty \) later. In analogy to the 
discussion in section \!\ref{KScoord}, we define 
\ber
u_{L} := \frac{a}{L} \;, v_{L} := \frac{M}{L} \,, \qquad
u_{R} := \frac{a}{R} \;, v_{R} := \frac{M}{R} 
\eer{uvdeff}
as new (small) parameters which will allow us to expand the integrands of
$I_{\rho}$ and $I_{z}$ \re{IrhoIz} in a (double) Taylor series 
expansion\footnote{For convenience, 
recall that a function $f(u,v)$ which is analytic about the point 
\( (u,v)=(0,0) \) can be expanded in a double Taylor series as
\[ f(u,v) = \sum_{m \geq 0 } \sum_{n \geq 0} \frac{1}{m!} \frac{1}{n!} 
\left( \frac{\pa^{m+n} f}{\pa u^{m} \, \pa v^{n}} \right)_{(0,0)} 
\, u^{m} \, v^{n} \,. \] }
to evaluate $I_{\rho}$ and $I_{z}$ \re{IrhoIz}. 

Thus, the expansion of the integrand of $I_{\rho}$ is
\[ \frac{4 L^2 \rho }{\left(L^2+\rho ^2\right)^{3/2}} v_{L} 
+\frac{4 L^3 \rho}{\left(L^2+\rho ^2\right)^2} \Big( 1+ \frac{L^2}
{L^2+\rho ^2} \Big) v_{L}^{2} \]
to cubic order terms. Integrating this expression term by term on \( 
\rho \in [0,R) \) and back-substituting $v_{L}$ later,
we obtain
\ber
I_{\rho} = \frac{1}{4} \lim _{\substack{L \to \infty \\ R \to \infty}} 
M \left(4-\frac{4 L}{\sqrt{L^2+R^2}} + \frac{3 M R^2}{L \left(L^2+R^2\right)}
+ \frac{M R^2 L}{ \left(L^2+R^2\right)^2} \right) \,.
\eer{Irhoara}
Finally, letting $L = k R$, we can see the effect of the chubbiness parameter $k$ 
on the result
\ber
I_{\rho} = \frac{1}{4} \lim_{R \to \infty} 
M \left(4-\frac{4 k}{\sqrt{k^2+1}} 
+ \frac{3 M}{R \, k \left(k^2+1\right)}
+ \frac{M k }{ R \left(k^2+1\right)^2} \right)
= M - M \frac{k}{\sqrt{k^2 +1}} \,.
\eer{Irhofin}
Similarly, the expansion of the integrand of $I_{z}$ is
\[ \frac{2 R^3 }{\left(R^2+z^2\right)^{3/2}} v_{R}
+ \frac{R^4}{\left(R^2+z^2\right)^2} \Big( \frac{5}{2} + 
\frac{2 z^2}{R^2+z^2} \Big) v_{R}^2 \]
to cubic order. Term by term integration of this expression on 
\( z \in (-L,L) \) and back-substitution of $v_{R}$ gives
\ber
I_{z} = \frac{1}{4} \lim _{\substack{R \to \infty \\ L \to \infty}} 
M \left(\frac{4 L}{\sqrt{L^2+R^2}} + \frac{M L^3 }{\left(L^2+R^2\right)^2}
+ \frac{2 M L}{L^2+R^2} + \frac{3 M}{R} \arctan{(L/R)} \right)  \,.
\eer{Izara}
Finally, with $L=k R$, we have
\ber
I_{z} = \frac{1}{4} \lim _{R \to \infty } 
M \left(\frac{4 k}{\sqrt{k^2+1}} + \frac{M k^3}{R \left(k^2+1\right)^2 }
+ \frac{2 M k }{R \left(k^2+1\right)} 
+ \frac{3 M}{R} \arctan{k} \right)
= M \frac{k}{\sqrt{k^2+1}} \,.
\eer{Izfin}
Substituting \re{Irhofin} and \re{Izfin} in \re{IrhoIz}, the total 
mass $Q(\bar \xi)$ is
\ber
Q(\bar{\xi}) = M \,,
\eer{MassWeyl}
as expected.

We proceed similarly for the Killing vector $\bar{\zeta}$ to calculate 
the angular momentum. We first write
\begin{align}
Q (\bar{\zeta}) & = J_{\rho} + J_{z} \,, \label{JrhoJz} \\
J_{\rho} & :=\frac{1}{4} \lim _{\substack{L \to \infty \\ R \to \infty}} 
\int_{0}^{R} d \rho \, \rho \, \Big( \ell^{tz} (\bar{\zeta}) \rvert_{z=L}
- \ell^{tz} (\bar{\zeta}) \rvert_{z=-L} \Big) \,, \nn \\
J_{z} & := \frac{1}{4} \lim _{\substack{R \to \infty \\ L \to \infty}}
\int_{-L}^{L} dz \, R \, \ell^{t\rho} (\bar{\zeta}) \rvert_{\rho=R} 
\nn \,.
\end{align}
Analogously, a series expansion of the integrand of $J_{\rho}$ gives
\[ \frac{6 L^3 \rho ^3}{\left(L^2+\rho ^2\right)^{5/2}} u_{L} v_{L}-
\frac{8 L^4 \rho ^3 }{\left(L^2+\rho ^2\right)^3} u_{L} v_{L}^2 \]
whereas that of $J_{z}$ is
\[ \frac{3 R^6}{\left(R^2+z^2\right)^{5/2}} u_{R} v_{R}-
\frac{4 R^7}{\left(R^2+z^2\right)^3} u_{R} v_{R}^2 \]
both to quartic order.
Integrating on \( \rho \in [0,R) \) and \( z \in (-L,L) \), respectively, and back-substituting, we get 
\begin{eqnarray*}
& & J_{\rho} = \frac{1}{4} \lim _{\substack{L \to \infty \\ R \to \infty}} 
2 a M \left( 2 - \frac{2 L}{\sqrt{ L^2+R^2 }} 
- \frac{L R^2} {\left(L^2+R^2\right)^{3/2}} 
-\frac{M R^4}{L \left(L^2+R^2\right)^2} \right) \, , \\
& & J_{z} = \frac{1}{4} \lim _{\substack{R \to \infty \\ L \to \infty}}
a M \left( \frac{4 L }{\sqrt{ L^2+R^2 }} 
+ \frac{2L R^2}{\left(L^2+R^2\right)^{3/2}}
-\frac{2 L M R^2}{\left(L^2+R^2\right)^2}
-\frac{3 L M}{L^2+R^2}
-\frac{3 M}{R} \arctan{(L/R)} \right)\,.
\end{eqnarray*}
Finally, letting $L= k R$, we have
\beq
\bald
\label{weylres}
J_{\rho} & = \frac{1}{4}  \lim _{ R \to \infty} 
2 a M \left( 2 - 2 \frac{k}{\sqrt{k^2+1}} 
- \frac{k}{\left( k^2+1 \right)^{3/2}}
- \frac{M}{R \, k \left(k^2 +1 \right)^{2} } \right) \\
&= a M \left(1- \frac{k}{\sqrt{k^2+1}} 
- \frac{1}{2} \frac{k}{\left( k^2+1 \right)^{3/2}} \right) \,, \\
J_{z} & = \frac{1}{4} \lim _{R \to \infty}
 a M \left( 4 \frac{k}{\sqrt{k^2+1}} 
+ \frac{2 k}{\left( k^2+1 \right)^{3/2}} 
-\frac{2 M k}{R \left( k^2+1 \right)^{2}}
-\frac{3 M k}{R \left( k^2+1 \right)}
-\frac{3 M}{R} \arctan{k}\right) \\
&= a M \left(\frac{k}{\sqrt{k^2+1}} 
+ \frac{1}{2} \frac{k}{\left( k^2+1 \right)^{3/2}} \right) \,.
\eald
\eeq
Substituting \re{weylres} in \re{JrhoJz}, the total angular momentum 
$Q(\bar \zeta)$  is 
\ber
Q(\bar{\zeta}) = a M \,,
\eer{AngMomWeyl}
as expected.

\section{Conclusions}\label{conc}
In this work, we have concisely reviewed the construction 
of asymptotic conserved gravitational charges using
background Killing vectors. One of the advantages of
this description is that the spacetime whose charges
are to be calculated does not need to have a Killing 
vector, rather it is enough if the spacetime it 
asymptotes to admits one itself.
Using this method, we have calculated the total 
mass/energy and the total angular momentum of the 
Kerr black hole with its timelike and rotational
Killing vectors in the Boyer-Lindquist, Kerr-Schild 
and Weyl canonical coordinates in sections \ref{KM}, 
\!\ref{KScoord}, \!\ref{Wcoord}, respectively, and shown
that they all agree, explicitly verifying the background 
gauge independence of this prescription. Even though
the calculations are harder for the Kerr-Schild 
and Weyl canonical coordinates, we have overcome 
the difficulties involved by introducing appropriate 
series expansions in suitably chosen ``small parameters".
We would like to emphasize that, to our knowledge, the 
presentations given in sections \!\ref{KScoord} and \!\ref{Wcoord} are original.

One of our main motivations was to examine how the gravitational
flux integration through ``the asymptotic boundary at infinity"
depended on the geometry of the integration surface. Obviously,
``the cube at infinity" for the Kerr-Schild form and ``the 
cylinder at infinity" for the Weyl canonical coordinates are not 
smooth geometries as for ``the sphere at infinity" for the 
Boyer-Lindquist coordinates. Hence, we were unable to determine 
the charges for the former via the shortcut of evaluating the 
limit of the relevant integrand at the spatial boundary 
and integrating the result on the boundary surface later, unlike 
the case for the Boyer-Lindquist coordinates. Moreover, 
for the calculation of the total angular momentum in
the case of the Kerr-Schild form with its boundary as 
a ``cube at infinity", note that the contribution of 
$J_{xy}$ \re{jxy} is less than \emph{half} of that of 
$J_{xz}$ \re{jxz}, which lets us deduce that the relevant 
gravitational flux is more abundant along the rotation axis 
compared to the other axes. In Weyl canonical coordinates, 
we introduced a chubbiness parameter $k$ while describing the 
spatial boundary. Although $k$ has an effect on the 
contribution of the relevant gravitational fluxes along 
different directions \re{weylres}, it is remarkable that 
the total charges remain independent of $k$.

For the case of the Kerr-Schild form, if we were to choose a
spatial boundary as a ``generic rectangular prism at infinity"
with three unequal side lengths, say \( (L_{x}, L_{y}, L_{z}) 
\), instead of one (equal) side length $L$ ``cube at infinity", 
we would expect to recover the correct calculations again by 
using three series expansions along the respective pairwise
side surfaces in the three ``small parameters" 
\( (a/L_{x}, a/L_{y}, a/L_{z}) \). In retrospect, 
we conjecture, at least heuristically, that provided the 
closed and simply connected spatial boundary is also piecewise
smooth, i.e. a finite union of smooth surfaces, the 
integration of the relevant gravitational flux, apart 
from obvious complications resulting in the pertinent 
calculations, does not depend on the geometry of the 
boundary.

Finally, as advocates of the use of background Killing 
isometries for the determination of gravitational charges, 
the details of our calculations should have at least some 
pedagogical value for those who want to appreciate the
delicate technicalities involved.

\bigskip

\noindent{\bf Acknowledgments}\\
It is a pleasure to thank Ulf Lindstr\"om for a careful reading of 
this manuscript.

\appendix
\section{The 2-form potential \texorpdfstring{$\ell$}{} in Weyl coordinates} \label{appendweyl}
For completeness sake, here we list the relevant non-vanishing components of \( \ell(\bar{\xi}) \) and \(\ell(\bar{\zeta})\)
that are needed in determining the gravitational charges in 
the Weyl canonical coordinates of section \!\ref{Wcoord}.
\begin{align*}
\ell^{t\rho}(\bar{\xi}) & = \frac{1}{8 \rho^3} \bigg(
\tfrac{\rho ^4 \left(\frac{a^2 (R_{-}-R_{+})^2}{M^2-a^2}+(2 M+R_{-}+R_{+})^2\right)}{R_{-}^3 R_{+}}+\tfrac{\rho ^4 \left(\frac{a^2 (R_{-}-R_{+})^2}{M^2-a^2}+(2 M+R_{-}+R_{+})^2\right)}{R_{-} R_{+}^3}
\\ & \hspace{-.8cm}
+ \rho^2 \left(\tfrac{\frac{a^2 (R_{-}-R_{+})^2}{M^2-a^2}+(2 M+R_{-}+R_{+})^2}{R_{-} R_{+}}-4\right)
-\tfrac{2 \rho ^4 \left(\left(a^2-M^2\right) (R_{-}+R_{+}) (2 M+R_{-}+R_{+})+a^2 (R_{-}-R_{+})^2\right)}{R_{-}^2 R_{+}^2 \left(a^2-M^2\right)} 
\\ & \hspace{-.8cm}
+\tfrac{8 \left(\rho ^2 \left(a^2 (R_{-}-R_{+})^2-\left(a^2-M^2\right) (2 M+R_{-}+R_{+})^2\right)^2-a^2 M^2 (2 M+R_{-}+R_{+})^2 \left(4 a^2-4 M^2+(R_{-}-R_{+})^2\right)^2\right)}{\left(a^2 (R_{-}-R_{+})^2-\left(a^2-M^2\right) (2 M+R_{-}+R_{+})^2\right) \left(4 a^2 \left(M^2-R_{-} R_{+}\right)-4 M^4+M^2 (R_{-}+R_{+})^2\right)}
\\ & \hspace{-.8cm}
+\tfrac{8 a^2 M^2 \rho ^2 (2 M+R_{-}+R_{+})^2 \left(4 a^2-4 M^2+(R_{-}-R_{+})^2\right)^2 \left(2 a^2 \left(R_{-}^2+R_{+}^2\right)-M^2 (R_{-}+R_{+})^2\right)}{R_{-} R_{+} \left(M^2 (2 M+R_{-}+R_{+})^2-4 a^2 (M+R_{-}) (M+R_{+})\right) \left(4 a^2 \left(M^2-R_{-} R_{+}\right)-4 M^4+M^2 (R_{-}+R_{+})^2\right)^2}
\\ & \hspace{-.8cm}
-\tfrac{8 \rho ^4 \left(M^2 (2 M+R_{-}+R_{+})^2-4 a^2 (M+R_{-}) (M+R_{+})\right) \left(2 a^2 \left(R_{-}^2+R_{+}^2\right)-M^2 (R_{-}+R_{+})^2\right)}{R_{-} R_{+} \left(4 a^2 \left(M^2-R_{-} R_{+}\right)-4 M^4+M^2 (R_{-}+R_{+})^2\right)^2}
\\ & \hspace{-.8cm}
+\tfrac{8 a^2 M^2 \rho ^2 (2 M+R_{-}+R_{+})^2 \left(4 a^2-4 M^2+(R_{-}-R_{+})^2\right)^2 \left(2 a^2 \left(M (R_{-}+R_{+})+R_{-}^2+R_{+}^2\right)-M^2 (R_{-}+R_{+}) (2 M+R_{-}+R_{+})\right)}{R_{-} R_{+} \left(a^2 (R_{-}-R_{+})^2-\left(a^2-M^2\right) (2 M+R_{-}+R_{+})^2\right)^2 \left(4 a^2 \left(M^2-R_{-} R_{+}\right)-4 M^4+M^2 (R_{-}+R_{+})^2\right)}
\\ & \hspace{-.8cm}
-\tfrac{8 \rho ^4 \left(M^2 (2 M+R_{-}+R_{+})^2-4 a^2 (M+R_{-}) (M+R_{+})\right)^2 \left(2 a^2 \left(M (R_{-}+R_{+})+R_{-}^2+R_{+}^2\right)-M^2 (R_{-}+R_{+}) (2 M+R_{-}+R_{+})\right)}{R_{-} R_{+} \left(a^2 (R_{-}-R_{+})^2-\left(a^2-M^2\right) (2 M+R_{-}+R_{+})^2\right)^2 \left(4 a^2 \left(M^2-R_{-} R_{+}\right)-4 M^4+M^2 (R_{-}+R_{+})^2\right)}
\\ & \hspace{-.8cm}
+\tfrac{8 a^2 M^2 \rho ^2 (2 M+R_{-}+R_{+}) \left(4 a^2-4 M^2+(R_{-}-R_{+})^2\right) \left(4 a^2 (R_{-}+R_{+})-4 M^2 (R_{-}+R_{+})-4 M (R_{-}-R_{+})^2-(R_{-}-R_{+})^2 (R_{-}+R_{+})\right)}{R_{-} R_{+} \left(M^2 (2 M+R_{-}+R_{+})^2-4 a^2 (M+R_{-}) (M+R_{+})\right) \left(4 a^2 \left(M^2-R_{-} R_{+}\right)-4 M^4+M^2 (R_{-}+R_{+})^2\right)}
\\ & \hspace{-1.2cm}
-\tfrac{8 \rho ^2 \left(4 a^2 \left(-\left(\rho ^2 \left(M (R_{-}+R_{+})+R_{-}^2+R_{+}^2\right)\right)-R_{-} R_{+} (M+R_{-}) (M+R_{+})\right)+M^2 (2 M+R_{-}+R_{+}) \left(R_{-} R_{+} (2 M+R_{-}+R_{+})+2 \rho ^2 (R_{-}+R_{+})\right)\right)}{R_{-} R_{+} \left(4 a^2 \left(M^2-R_{-} R_{+}\right)-4 M^4+M^2 (R_{-}+R_{+})^2\right)}
\\ & \hspace{-0.8cm}
+4 \left(\rho ^2-\tfrac{\rho ^2 \left(a^2 (R_{-}-R_{+})^2-\left(a^2-M^2\right) (2 M+R_{-}+R_{+})^2\right)^2-a^2 M^2 (2 M+R_{-}+R_{+})^2 \left(4 a^2-4 M^2+(R_{-}-R_{+})^2\right)^2}{\left(a^2 (R_{-}-R_{+})^2-\left(a^2-M^2\right) (2 M+R_{-}+R_{+})^2\right) \left(4 a^2 \left(M^2-R_{-} R_{+}\right)-4 M^4+M^2 (R_{-}+R_{+})^2\right)}\right) \bigg) \,,
\\ \vspace{1cm}
\ell^{tz}(\bar{\xi}) & = \frac{1}{8} \bigg(
\tfrac{\left(z-\sqrt{M^2-a^2}\right) \left(\frac{a^2 (R_{-}-R_{+})^2}{M^2-a^2}+(2 M+R_{-}+R_{+})^2\right)}{R_{-}^3 R_{+}}
+\tfrac{\left(\sqrt{M^2-a^2}+z\right) \left(\frac{a^2 (R_{-}-R_{+})^2}{M^2-a^2}+(2 M+R_{-}+R_{+})^2\right)}{R_{-} R_{+}^3}
\\ & \hspace{-.8cm}
-\tfrac{\frac{2 a^2 (R_{-}-R_{+}) \left(\frac{z-\sqrt{M^2-a^2}}{R_{-}}-\frac{\sqrt{M^2-a^2}+z}{R_{+}}\right)}{M^2-a^2}+2 (2 M+R_{-}+R_{+}) \left(\frac{z-\sqrt{M^2-a^2}}{R_{-}}+\frac{\sqrt{M^2-a^2}+z}{R_{+}}\right)}{R_{-} R_{+}}
\\ & \hspace{-.8cm}
+\tfrac{8\left(\frac{R_{+} \left(2 a^2-M^2\right) \left(\sqrt{M^2-a^2}-z\right)}{R_{-}}-\frac{R_{-} \left(2 a^2-M^2\right) \left(\sqrt{M^2-a^2}+z\right)}{R_{+}}+2 M^2 z\right)}{\rho ^2 \left(a^2 (R_{-}-R_{+})^2-\left(a^2-M^2\right) (2 M+R_{-}+R_{+})^2\right) \left(4 a^2 \left(M^2-R_{-} R_{+}\right)-4 M^4+M^2 (R_{-}+R_{+})^2\right)^2} \times
\\ & \hspace{-.5cm}
\scalebox{0.8}{$\left( \rho ^2 \left(a^2 (R_{-}-R_{+})^2-\left(a^2-M^2\right) (2 M+R_{-}+R_{+})^2\right)^2-a^2 M^2 (2 M+R_{-}+R_{+})^2 \left(4 a^2-4 M^2+(R_{-}-R_{+})^2\right)^2 \right)$}
\\ & \hspace{-1.2cm}
-\tfrac{8 a^2 (R_{-}-R_{+}) \left(R_{-} \left(\sqrt{M^2-a^2}+z\right)+R_{+} \left(\sqrt{M^2-a^2}-z\right)\right)+\left(a^2-M^2\right) (2 M+R_{-}+R_{+}) \left(R_{-} \left(\sqrt{M^2-a^2}+z\right)+R_{+} \left(z-\sqrt{M^2-a^2}\right)\right)}{\rho ^2 R_{-} R_{+} \left(a^2 (R_{-}-R_{+})^2-\left(a^2-M^2\right) (2 M+R_{-}+R_{+})^2\right)^2 \left(4 a^2 \left(M^2-R_{-} R_{+}\right)-4 M^4+M^2 (R_{-}+R_{+})^2\right)} \times
\\ & \hspace{-.5cm}
\scalebox{0.8}{$\left( \rho ^2 \left(a^2 (R_{-}-R_{+})^2-\left(a^2-M^2\right) (2 M+R_{-}+R_{+})^2\right)^2-a^2 M^2 (2 M+R_{-}+R_{+})^2 \left(4 a^2-4 M^2+(R_{-}-R_{+})^2\right)^2 \right)$}
\\ & \hspace{-.8cm}
+\tfrac{16}{\rho ^2 R_{-} R_{+} \left(a^2 (R_{-}-R_{+})^2-\left(a^2-M^2\right) (2 M+R_{-}+R_{+})^2\right) \left(4 a^2 \left(M^2-R_{-} R_{+}\right)-4 M^4+M^2 (R_{-}+R_{+})^2\right)} \times
\\ & \hspace{-1cm}
\scalebox{0.8}{$\Big(
-a^2 M^2 (R_{-}-R_{+}) (2 M+R_{-}+R_{+})^2 \left(4 a^2-4 M^2+(R_{-}-R_{+})^2\right) \left(R_{-} \left(\sqrt{M^2-a^2}+z\right)+R_{+} \left(\sqrt{M^2-a^2}-z\right)\right)
\Big.$}
\\ & \hspace{-1cm}
\scalebox{0.8}{$+\frac{1}{2} a^2 M^2 (2 M+R_{-}+R_{+}) \left(4 a^2-4 M^2+(R_{-}-R_{+})^2\right)^2 \left(R_{-} \left(\sqrt{M^2-a^2}+z\right)+R_{+} \left(z-\sqrt{M^2-a^2}\right)\right)$}
\\ & \hspace{-1cm}
\scalebox{0.7}{$+\rho^2 \left( a^2 (R_{-}-R_{+})^2-\left(a^2-M^2\right) (2 M+R_{-}+R_{+})^2 \right) \times $}
\\ & \hspace{-.6cm}
\scalebox{0.7}{$\Big. \left( a^2 (R_{-}-R_{+}) \left(R_{-} \left(\sqrt{M^2-a^2}+z\right)+R_{+} \left(\sqrt{M^2-a^2}-z\right)\right)+\left(a^2-M^2\right) (2 M+R_{-}+R_{+}) \left(R_{-} \left(\sqrt{M^2-a^2}+z\right)+R_{+} \left(z-\sqrt{M^2-a^2}\right)\right) \right)
\Big)$}
\bigg) \,, 
\end{align*}

\begin{align*}
\ell^{t\rho}(\bar{\zeta}) & =
\tfrac{a M}{2 \rho  R_{-} R_{+} 
\left(M^2 (2 M+R_{-}+R_{+})^2-4 a^2 (M+R_{-}) (M+R_{+})\right)^2} \times
\\ & \hspace{-.5cm} \bigg(
\scalebox{0.8}{$
-2 R_{-} R_{+} (2 M+R_{-}+R_{+}) \left(4 a^2-4 M^2+(R_{-}-R_{+})^2\right) \left(M^2 (2 M+R_{-}+R_{+})^2-4 a^2 (M+R_{-}) (M+R_{+})\right) $}
\\ & \hspace{-.5cm}
\scalebox{0.8}{$
+ \rho ^2 (R_{-}+R_{+}) +\left(4 a^2-4 M^2+(R_{-}-R_{+})^2\right) \left(M^2 (2 M+R_{-}+R_{+})^2-4 a^2 (M+R_{-}) (M+R_{+})\right) $}
\\ & \hspace{-.5cm}
\scalebox{0.8}{$
-2 \rho ^2 (R_{-}-R_{+})^2 (2 M+R_{-}+R_{+}) \left(M^2 (2 M+R_{-}+R_{+})^2-4 a^2 (M+R_{-}) (M+R_{+})\right) $}
\\ & \hspace{-.5cm}
\scalebox{0.8}{$
+2 \rho ^2 (2 M+R_{-}+R_{+}) \left(4 a^2-4 M^2+(R_{-}-R_{+})^2\right) \left(2 a^2 \left(M (R_{-}+R_{+})+R_{-}^2+R_{+}^2\right)-M^2 (R_{-}+R_{+}) (2 M+R_{-}+R_{+})\right) $}
 \bigg) \,,
\\
\ell^{tz}(\bar{\zeta}) & =
\tfrac{a M}{2 R_{-} R_{+} 
\left(M^2 (2 M+R_{-}+R_{+})^2-4 a^2 (M+R_{-}) (M+R_{+})\right)^2} \times
\\ & \hspace{-.5cm} \bigg(
\scalebox{0.8}{$
-2 (R_{-}-R_{+}) (2 M+R_{-}+R_{+}) \left(M^2 (2 M+R_{-}+R_{+})^2-4 a^2 (M+R_{-}) (M+R_{+})\right) \left(R_{-} \left(\sqrt{M^2-a^2}+z\right)+R_{+} \left(\sqrt{M^2-a^2}-z\right)\right)$} 
\\ & \hspace{-.5cm}
\scalebox{0.8}{$
+\left(4 a^2-4 M^2+(R_{-}-R_{+})^2\right) \left(M^2 (2 M+R_{-}+R_{+})^2-4 a^2 (M+R_{-}) (M+R_{+})\right) \left(R_{-} \left(\sqrt{M^2-a^2}+z\right)+R_{+} \left(z-\sqrt{M^2-a^2}\right)\right)$}
\\ & \hspace{-.5cm}
\scalebox{0.7}{$
-2 (2 M+R_{-}+R_{+}) \left(4 a^2-4 M^2+(R_{-}-R_{+})^2\right) 
\times $}
\\ & 
\scalebox{0.7}{$
\left(R_{+} \left(\sqrt{M^2-a^2}-z\right) \left(2 a^2 (M+R_{+})-M^2 (2 M+R_{-}+R_{+})\right)-R_{-} \left(\sqrt{M^2-a^2}+z\right) \left(2 a^2 (M+R_{-})-M^2 (2 M+R_{-}+R_{+})\right)\right)$}
\bigg) \,.
\end{align*}

\end{document}